\documentclass[ aps,
                prl,
                twocolumn
                ]{revtex4-2}
\usepackage{graphicx}
\usepackage{ulem}
\usepackage[dvipsnames]{xcolor}
\usepackage{amsmath}
\usepackage{amssymb}
\usepackage[colorlinks=true,urlcolor=blue,citecolor=blue,linkcolor=magenta]{hyperref}
\usepackage{braket}
\usepackage{physics}

\begin{document}
\title{Remote Phase Sensing by Coherent Single Photon Addition}

\author{Nicola~Biagi$^{1,2}$, Saverio~Francesconi$^{1,2}$, Manuel~Gessner$^{3}$, Marco~Bellini$^{1,2^*}$, and Alessandro~Zavatta$^{1,2^*}$}

\affiliation{$^{1}$Istituto Nazionale di Ottica (CNR-INO), L.go E. Fermi 6, 50125 Florence, Italy\\
$^{2}$LENS and Department of Physics $\&$ Astronomy, University of Firenze, 50019 Sesto Fiorentino, Florence, Italy\\$^{3}$Laboratoire Kastler Brossel, ENS-Universit\'{e} PSL, CNRS, Sorbonne Universit\'{e}, Coll\`{e}ge de France, 24 Rue Lhomond, 75005, Paris, France}

\begin{abstract}
We propose a remote phase sensing scheme inspired by the high sensitivity of the entanglement produced by coherent multimode photon addition on the phase set in the remote heralding apparatus. By exploring the case of delocalized photon addition over two modes containing identical coherent states, we derive the optimal observable to perform remote phase estimation from heralded quadrature measurements. The technique is experimentally tested with calibration measurements and then used for estimating a remote phase with a sensitivity that is found to scale with the intensity of the (local) coherent states, which never interacted with the sample.
\end{abstract}
\date{\today}
\maketitle

\bigskip
\bigskip

\begin{figure}[ht]
 		\centering
 		\includegraphics[width=0.95\columnwidth]{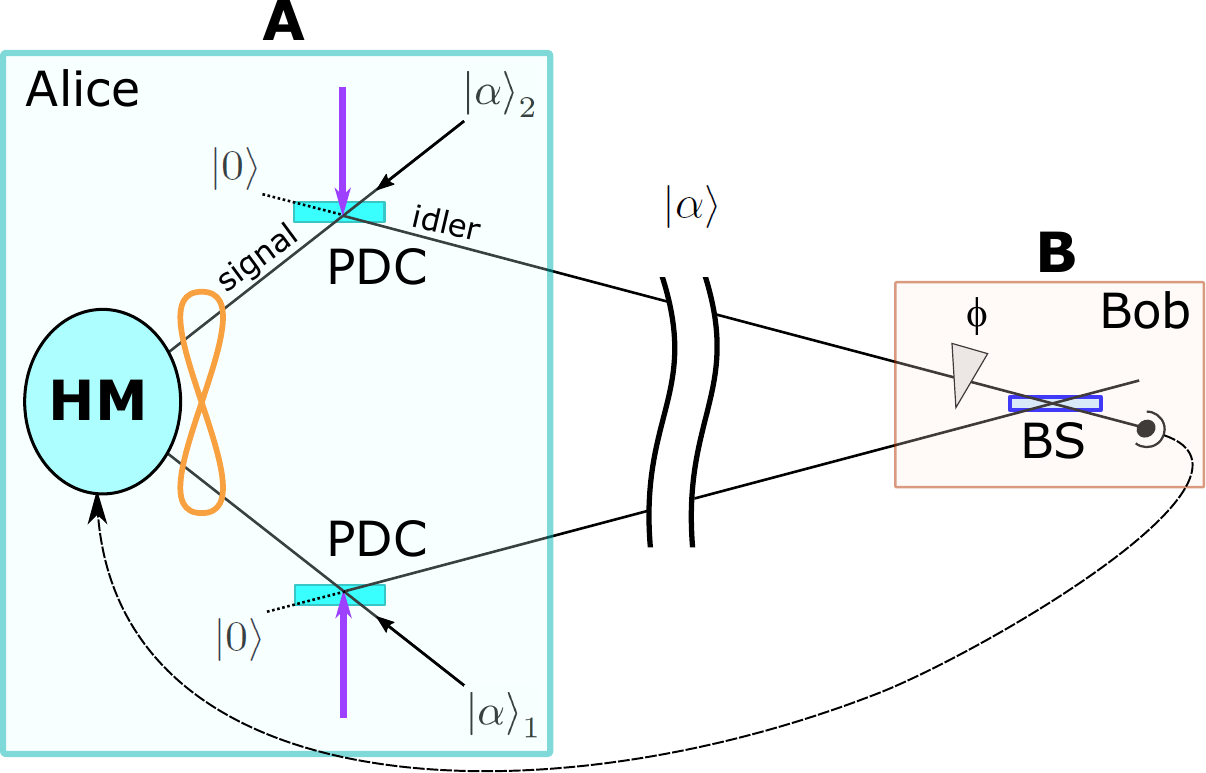}
 		\caption{
 		        General scheme of the remote phase sensing experiment. HM represents some heralded measurement on the two signal modes in Alice's lab triggered by the remote detector in Bob's lab.
 			}
 		\label{fig:ExpSetup}
 	\end{figure}

Although the strong dependence of entanglement on external interactions is often detrimental to its experimental generation and manipulation, especially for multi-photon states, one may exploit this feature and the nonlocal character of entanglement to gain sensitive information about remote physical systems.
Suppose Alice and Bob are two researchers located in their distant laboratories, A and B, willing to measure the unknown phase introduced by a sample in B. In her lab, Alice may inject two identical coherent states (CS) $|\alpha\rangle$ in the signal modes of two parametric down-conversion (PDC) crystals (see Fig.~\ref{fig:ExpSetup}). The two corresponding idler modes, initially empty, are then sent to the remote location B, where Bob can interfere them on a beam-splitter (BS) after passing one of them through the sample under investigation, which may introduce an additional phase shift. 

The detection of a single photon after Bob's BS heralds the delocalized addition of a single photon \cite{entropy21} and the generation of entanglement between the two signal modes in Alice's lab, yielding the state
\begin{align}\label{eq:state}
|\Psi(\phi)\rangle=\mathcal{N}(\hat{a}_1^{\dagger}+e^{i\phi}\hat{a}_2^{\dagger})\ket{\alpha}_1 \ket{\alpha}_2,
\end{align}
where $\mathcal{N}=1/\sqrt{2(1+(1+\cos\phi)|\alpha|^2)}.$
The phase between the idler modes in B thus determines the phase $\phi$ of the operator superposition, and the heralded state in A changes correspondingly. Also the degree of entanglement between the two signal modes in A changes very rapidly (depending on the amplitude of the CS) with $\phi$ \cite{BCBZ20,suppl}. One could thus perform heralded measurements (HM) of the entanglement of the state in A to probe, remotely, the phase in B.

However, measuring entanglement normally requires a non-trivial full tomographic reconstruction of the state \cite{Lee00,Vidal02} and may not be the most convenient measurement scheme that one can think of. Alternatively, Alice may recombine her two signal modes on another BS, forming a local interferometer in A that is connected to Bob's distant interferometer by the delocalized photon addition operation. Remote phase changes in B would thus be measurable by heralded interferometric measurements in A. 

Besides its fundamental interest, this kind of nonlocal interferometry \cite{Franson07} presents several practical advantages over conventional schemes in sensing applications: (I) first of all, the sample could be placed in a remote location that is not easily accessible and might be far away from the fully equipped Alice's lab; (II) the scheme is very robust against losses in the idler channel since their effect would only decrease the measurement rate, without affecting the measurement sensitivity; (III) different wavelengths could also be used for signal and idler modes (by using non-frequency-degenerate PDC) to meet experimental constrains (e.g. transmission over long distances, in air or fibers, best coupling to the sample, best performance of the detectors, etc.); (IV) the practical realization of this interferometric scheme does not even require real-time communication between Alice and Bob, since the protocol can be implemented off-line by post-selection of the tagged events. 

\begin{figure}[ht]
 		\centering
 		\includegraphics[width=0.95\columnwidth]{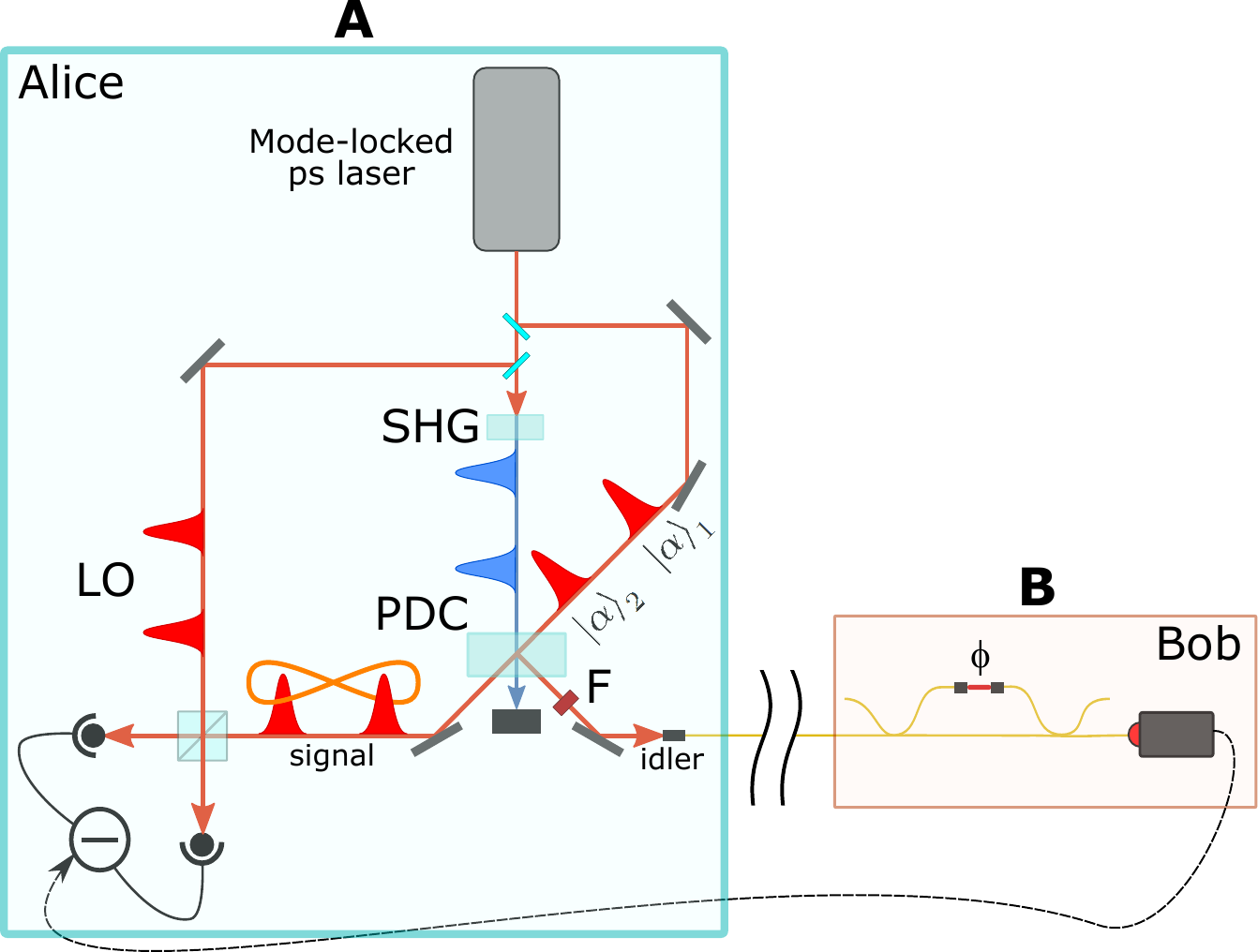}
 		\caption{
 		        Scheme of the actual remote phase sensing experiment. The main light source is a mode-locked Ti:Sa laser emitting 1.5 ps-long pulses at 786 nm with a repetition rate of 81 MHz. It provides the local oscillator (LO) pulses for homodyne detection, the pump pulses for the PDC process after a frequency-doubling stage (SHG), and the coherent states that seed the PDC crystal. Bob's interferometer is stabilized by means of a feedback loop (not shown in the figure) using a counter-propagating laser beam injected in the unused output port and the same beam is also employed to classically estimate the phase $\phi$ by measuring interference fringes.
 			}
 		\label{fig:ExpSetupTemporalModes}
 \end{figure}

The actual experimental setup used to implement this technique relies on temporal modes \cite{remoteprep,hybrid} instead of spatial ones, and it is shown in  Fig.~\ref{fig:ExpSetupTemporalModes}. 

The two photon additions \cite{progopt} in Alice's lab are obtained with a single PDC device, a type-I beta-barium borate (BBO) crystal, operating onto two co-propagating consecutive wavepacket temporal modes. These are then analyzed via a single time-multiplexed homodyne detector, which implements the generic heralded measurement of Fig.~\ref{fig:ExpSetup}. The idler photons produced by the PDC process are spectrally and spatially filtered (F) and then sent along the same spatial mode to Bob's side, where a fiber-based Mach-Zehnder interferometer, having the arms unbalanced by exactly the time difference between these temporal modes, coherently superposes the two. A single-photon detector, placed in one of the output ports of the interferometer, heralds the delocalized single-photon addition.
The phase $\phi$ of the state superposition is adjusted by finely varying an air-gap length, which emulates the effect of the sample, in one of the interferometer arms.

Such a temporal-mode version has already been demonstrated to work as expected and to generate the desired states with measurements of entanglement and discorrelation \cite{BCBZ20, Discor}. However, using this scheme might offer even more advantages for remote interferometry applications.
Indeed, since a single optical path (e.g. a single-mode fiber) is used to connect Alice's and Bob's labs, the idler modes suffer approximately the same phase noise until they reach Bob's interferometer. For this reason, Bob only has to compensate the local fluctuations of his interferometer. 
Moreover, on Alice's site, the phase between the two consecutive local oscillator pulses coming from the same mode-locked laser is much simpler to control and stabilize in this case, compared to the scheme with separated spatial modes, and the same advantage holds for the two identical coherent states injected in the signal PDC modes.

Since Alice can straightforwardly perform homodyne measurements on both her signal modes, let us now see what is the most practical way to estimate the remote phase $\phi$ in B from quadrature measurements in A. Specifically, we consider the measurement of a single observable $\hat{X}=c_{x1}\hat{x}_1+c_{p1}\hat{p}_1+c_{x2}\hat{x}_2+c_{p2}\hat{p}_2$ that is constructed from the quadratures of both modes in Alice's lab. The variation of this observable's expectation value as a function of the phase $\phi$ is determined during calibration and can be predicted theoretically. To estimate the value of $\phi$, we compare the sample average of $\mu$ measurements of $\hat{X}$ to this calibration curve. In the limit of many repeated measurements, $\mu\gg 1$, this yields a phase estimation error of:
\begin{align}\label{eq:phierror}
    (\Delta\phi_{\mathrm{est}})^2=\frac{1}{\mu}\frac{(\Delta \hat{X})^2}{\left|\partial_{\phi}\langle\hat{X}\rangle\right|^2} = \frac{1}{\mu\,S},
\end{align}
where $S$ is the measurement sensitivity. In order to minimize the estimation error, we optimize the choice of the real coefficients $c_{x1},c_{p1},c_{x2},c_{p2}$ that determine the observable $\hat{X}$. The optimal observable is determined analytically from theoretical predictions of the first and second moments of the state~(\ref{eq:state}) using the method introduced in Ref.~\cite{GessnerPRL2019} (see \cite{suppl} for details). The highest sensitivity is found in the vicinity of $\phi=\pi$, which is where we will operate our interferometer in the following. We obtain a maximal sensitivity of
\begin{align}\label{eq:maxsensitivity}
    \max_{\hat{X}}\, S=\frac{2}{3}|\alpha|^2,
\end{align}
which is attained by the optimal observable $\hat{X}_{\mathrm{opt}}=\frac{\mathrm{Im}(\alpha)}{|\alpha|}(\hat{x}_2-\hat{x}_1)+\frac{\mathrm{Re}(\alpha)}{|\alpha|}(\hat{p}_1-\hat{p}_2)$. For real $\alpha$, this identifies $\hat{X}_{\mathrm{opt}}=\hat{p}_1-\hat{p}_2$ as the optimal observable for our scheme.

To gauge the quality of this simple observable, we compare it to the ultimate quantum limit on the estimation error, which is determined by the quantum Cram\'er-Rao bound as $(\Delta\phi_{\mathrm{est}})^2\geq(\mu F_Q[\Psi(\phi,\alpha)])^{-1}$~\cite{Paris2009,GiovannettiNatPhoton2011,PezzeSmerzi2014}. The quantum Fisher information $F_Q[\Psi(\phi,\alpha)]$ quantifies the maximal sensitivity of the state $\Psi(\phi,\alpha)$, which may be attained by an unconstrained optimization over the measurement observable~\cite{BraunsteinCavesPRL1994}, i.e., by considering also measurements beyond homodyne techniques. Here, we obtain $F_Q[\Psi(\pi,\alpha)]=2|\alpha|^2+1$, which corresponds to the total mean number of photons in the heralded state~(\ref{eq:state}). An optimal quadrature measurement thus achieves the best possible sensitivity scaling with the mean number of photons in Alice's modes, whereas a quantum-optimal measurement would be able to improve the pre-factor.

In order to validate the proposed technique for phase estimation, we first perform the calibration procedure, verifying the expected behaviour \begin{equation}\label{eq_meanoperator}
    \expval*{\hat{X}}=-\frac{\alpha \sin(\phi)}{1+\alpha^2+\alpha^2 \cos(\phi)}
\end{equation} 
for the mean value of the observable $\hat{X}=(\hat{p}_1-\hat{p}_2)$ evaluated on the quantum state $\ket{\Psi(\phi,\alpha)}$ of Eq.~(\ref{eq:state}). Here and in the following, we consider real values for the coherent state amplitudes $\alpha$. Then, we carry out a phase estimation, evaluating the variance of the estimator and thus the sensitivity of the technique.

\begin{figure}[t]
  \includegraphics[width=0.75\columnwidth]{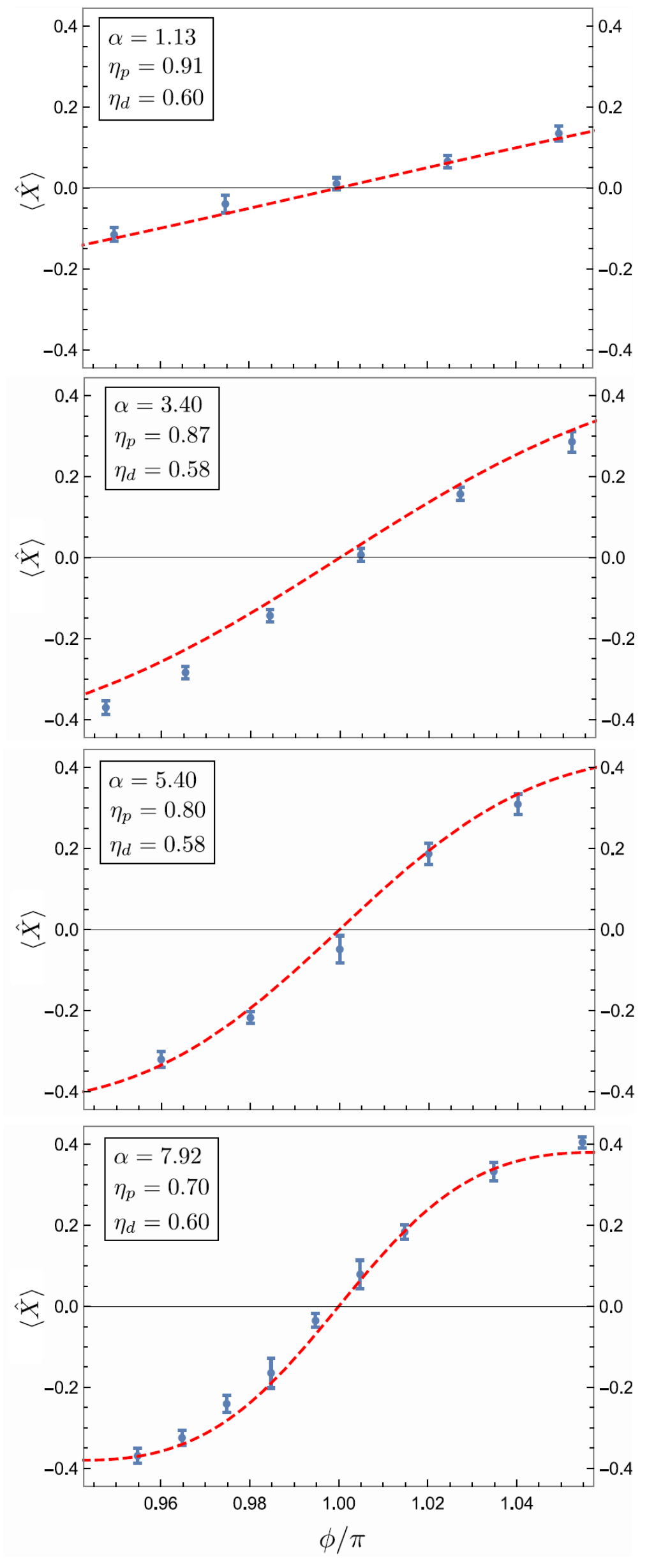}
  \caption{Calibration curves: experimental (blue dots) and theoretical (red dashed curve) expectation values $\expval*{\hat{X}}=\expval{\hat{p}_1-\hat{p}_2}$ as a function of the phase $\phi$ of the state superposition for different values of $\alpha$. Errors are evaluated by dividing the quadrature data into sets of 5000 values each, resulting in 10 samples for each value of $\phi$. The error bars are the standard deviation of these 10 values of $\expval*{\hat{X}}$. The theoretical curves are calculated by employing the detection efficiency of our setup and taking into account the nonperfect visibility of the Mach-Zehnder interferometer, which degrades the preparation efficiency. Their values are reported in the top-left of each graph \cite{suppl}.}
  \label{fig:CalibrationCurves}
\end{figure}

The calibration works as follows. First of all, the phase of Bob's interferometer $\phi$ is evaluated classically employing a bright counter-propagating laser beam. Then, while the phase is scanned around $\pi$, we perform homodyne measurements of the state in the signal mode, for several seed amplitudes $\alpha$. For each value of the phase, about $50000$ two-mode quadrature measurements are performed, retrieving the mean value and the variance of the observable $\hat{X}$.
The results are shown in Fig.~\ref{fig:CalibrationCurves} for four values of $\alpha$ (1.13, 3.40, 5.40, 7.92). In the vicinity of $\phi=\pi$, the experimental points show an evident linear dependence $\expval*{\hat{X}} \approx \alpha (\phi-\pi)$, with a slope increasing with $\alpha$, as expected from Eq.~\ref{eq_meanoperator}. In particular, they are in very good agreement with the theoretical curves calculated by also taking into account the independently-determined preparation and detection efficiencies, $\eta_p$ and $\eta_d$, which are reported on the top-left corner of each graph \cite{suppl}. Therefore, the theoretical curves can be employed as calibration curves of the phase estimation technique.

\begin{figure}[ht]
  \includegraphics[width=0.75\columnwidth]{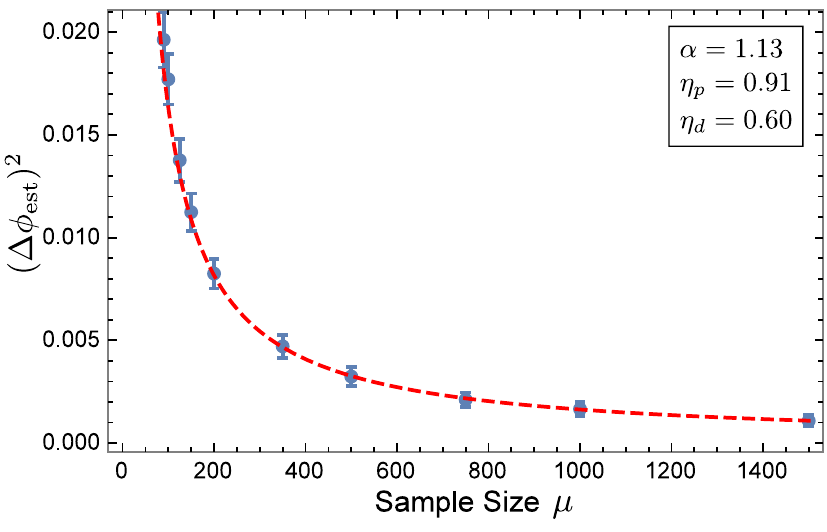}
  \caption{Variance $(\Delta\phi_\mathrm{est})^2$ of the estimator of $\phi$ for $\alpha=1.13$ as a function of the sample size $\mu$: experimental points in blue and theoretical curve in red. Statistical errors are evaluated with a bootstrap method using resampled quadrature data sets. For each data set we evaluate the estimator of $\phi$ and the estimation error $(\Delta\phi_\mathrm{est})^2$. Error bars in the plot are calculated as the standard deviation of the latter.}
\label{fig:PhaseEstimationErrorAlpha1}
\end{figure}

We then perform a phase estimation in the case $\alpha=1.13$, with the interferometer phase set to an "unknown" value near $\pi$. We take the sample average of $\mu$ measurements of $\expval*{\hat{X}}$ and, by inverting the theoretical calibration curve (Fig.~\ref{fig:CalibrationCurves}), we evaluate the estimator of $\phi$ and its variance $(\Delta\phi_\mathrm{est})^2$, which quantifies the estimation error. 
The latter is reported with blue dots in Fig.~\ref{fig:PhaseEstimationErrorAlpha1} as a function of the sample size $\mu$, compared to the theoretical model (red dashed curve) which takes into account the experimental imperfections.
Asymptotically, in the limit of large $\mu$, the estimation error $(\Delta\phi_\mathrm{est})^2$ converges to Eq.~(\ref{eq:phierror}) with the sensitivity $S$ ideally given by~(\ref{eq:maxsensitivity}), but modified in order to consider the experimental inefficiencies \cite{suppl}. 

\begin{figure}[b]
  \includegraphics[width=0.75\columnwidth]{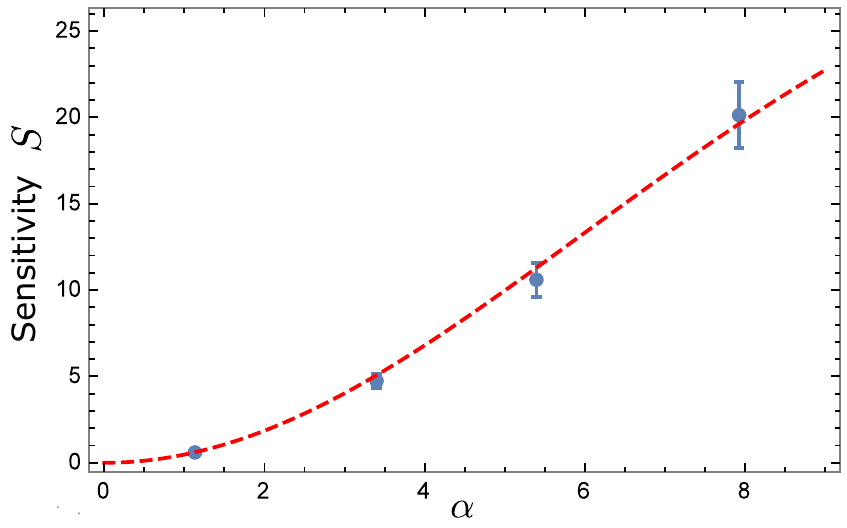}
  \caption{Sensitivity of the remote phase estimation as a function of the (local) seed amplitude $\alpha$: experimental data (blue points) and expected behavior (red dashed curve). The latter is calculated by taking into account the $\alpha$ dependence of the preparation efficiency $\eta_p$ and the mean value of the detection efficiency of the four experimental points $\bar{\eta}_d=0.59$. Statistical errors are evaluated with a bootstrap method using resampled data sets with $\mu=200$.}
   \label{fig:SenstivityExp}
\end{figure}

The sensitivity of the phase estimation can be experimentally evaluated as the product of the sample size and the variance of the estimator, $S^{-1}=\mu \cdot (\Delta\phi_\mathrm{est})^2 $, in the limit $\mu\gg 1$.
By repeating the same phase estimation procedure for several values of the seed amplitude, we then aim at verifying the expected dependence of the sensitivity on $\alpha$, as ideally given by Eq. (\ref{eq:maxsensitivity}).
The experimentally obtained values of sensitivity are presented in Fig.~\ref{fig:SenstivityExp}. They closely follow the theoretical curve calculated by taking into account the experimental imperfections and, in particular, the fact that the state preparation efficiency decreases when the seed amplitude $\alpha$ increases \cite{suppl}. The resulting dependence of the measured sensitivity on $\alpha$ is thus slightly less than quadratic but, nevertheless, it validates the proposed technique for remote phase estimation.

In the ideal case, the sensitivity of the method is directly proportional to the mean number of photons in the seed coherent states in Alice's lab, while the sample, placed in the idler path at Bob's lab, is only crossed by the much weaker PDC emission, which is reduced by a factor $g^2$ proportional to the parametric gain (in our case $g^2 \approx 10^{-6}$).

Although it would appear possible to arbitrarily improve the phase estimation sensitivity by increasing the intensity of the coherent states in A without harming a possibly delicate sample in B, the non-deterministic character of the measurement procedure implies a low heralding rate. In the vicinity of $\phi=\pi$, the measurement rate also reduces to the spontaneous PDC rate of $g^2$ events per pulse and thus compensates for possible ``quantum'' advantages when the actual total mean number of photons crossing the sample is considered.

In conclusion, we have demonstrated the possibility of performing remote phase sensing by coherently adding a single photon over two separated field modes. In the particular version employing collinear temporal modes that we used for the experimental implementation, the scheme is reminiscent of a Franson's interferometer for the analysis of time-frequency entanglement \cite{Franson89,Ou90,Brendel99}. As in that case, the results of coincidence measurements depend on both the phases in the remote locations A and B.
Here, thanks to the stimulated photon addition process onto populated signal modes, the entanglement of the heralded state in A presents an enhanced sensitivity on the remote phase $\phi$ that allows us to perform phase measurements on the sample in B with a sensitivity scaling with the intensity of the seed coherent states in A. Actually, the phase-changing sample does not even need to be in Bob's lab and could be located anywhere along the path of one of the idler modes (or even distributed in the white zone between the A and B blocks of Fig.~\ref{fig:ExpSetup} and Fig.~\ref{fig:ExpSetupTemporalModes}). In such a case, Bob's only task would be to report the clicks of his detector to Alice and she might be able to perform phase measurements on the distant sample without Bob ever acquiring any information about it. For example, Alice might add a random phase to one of her seed coherent states and perform her heralded measurements only when the two coherent states are in phase. Without this extra information, Bob's clicks would be totally meaningless to him.
Although some more investigations and a rigorous security analysis are certainly needed, we expect that the approach presented here might thus find interesting applications also in particular secure communication tasks.

\smallskip
\section{Acknowledgments}
N.B., S.F., M.B., and A.Z. gratefully acknowledge the support of the EU under the ERA-NET QuantERA project ``ShoQC'' (grant No. 731473) and the FET Flagship on Quantum Technologies project "Qombs" (grant no. 820419). M.G. acknowledges funding by the LabEx ENS-ICFP: ANR-10-LABX-0010 / ANR-10-IDEX-0001-02 PSL.

\bigskip
$^*$Corresponding authors.\\marco.bellini@ino.cnr.it\\alessandro.zavatta@ino.cnr.it

\pagebreak

\begin{widetext}
\section{Supplemental Material}

\subsection{Entanglement of the heralded state}
The amount of entanglement present in a quantum state can be quantified evaluating the Negativity under Partial Transposition ($NPT$) of its density matrix \cite{Horodecki96,Peres96,Lee00}. A value of $NPT = 1$ identifies a maximally entangled state, while a completely separable state has $NPT = 0$. Considering the state in Eq.(1) of the main paper, the $NPT$ becomes:
\begin{equation}
	NPT(\phi)=\frac{1}{1+|\alpha |^2(1+ \cos(\phi))}.
\end{equation}
%\mg{For general, complex $\alpha$ same formula with $|\alpha|$ instead?}{}
As shown in Fig.~\ref{fig:NPT}, the $NPT$ reaches 1 only for $\phi=\pi$, while the minimum corresponds to $\phi=0$. The slope of this curve increases with the amplitude of the coherent states ($\alpha$), making this parameter very sensitive to phase changes around $\pi$. This kind of behaviour, already observed in \cite{BCBZ20}, motivated this study.
\begin{figure}[h]
	\includegraphics[width=0.6\columnwidth]{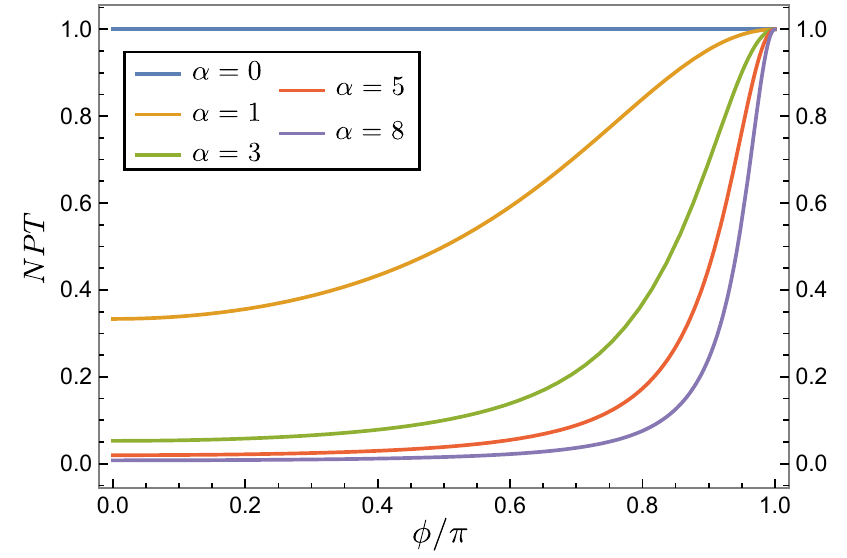}
	\caption{$NPT$ behaviour, calculated for the state in Eq.(1) of the main paper, as a function of the superposition phase $\phi$, for different values of $\alpha$.}
	\label{fig:NPT}
\end{figure}

\subsection{Optimization over the measurement observable}
To identify the optimal homodyne measurement observable, we maximize the sensitivity $S$ by varying the coefficients $\mathbf{c}=(c_{x1},c_{p1},c_{x2},c_{p2})^T$ in the expansion of the observable $\hat{X}=c_{x1}\hat{x}_1+c_{p1}\hat{p}_1+c_{x2}\hat{x}_2+c_{p2}\hat{p}_2$. It was shown in Ref.~\cite{GessnerPRL2019} that
\begin{align}\label{eq:maxbound}
	\max_{\mathbf{c}}\frac{\left|\partial_{\phi}\langle \hat{X}\rangle\right|^2}{(\Delta\hat{X})^2}=\mathbf{D}^T\boldsymbol{\Gamma}^{-1}\mathbf{D},
\end{align}
where
\begin{align}\label{eq:covA}
	(\boldsymbol{\Gamma})_{ij}=\frac{1}{2}\left(\langle \hat{r}_i\hat{r}_j\rangle+\langle \hat{r}_j\hat{r}_i\rangle\right)-\langle \hat{r}_i\rangle\langle \hat{r}_j\rangle
\end{align}
is the $4\times 4$ quadrature covariance matrix and the $\hat{r}_i$ are the elements of the quadrature vector $\hat{\mathbf{r}}=(\hat{x}_1,\hat{p}_1,\hat{x}_2,\hat{p}_2)^T$. Moreover,
\begin{align}\label{eq:DA}
	\mathbf{D}=(\partial_{\phi}\langle \hat{x}_1\rangle,\partial_{\phi}\langle \hat{p}_1\rangle,\partial_{\phi}\langle \hat{x}_2\rangle,\partial_{\phi}\langle \hat{p}_2\rangle)^T,
\end{align}
describes the variation of each quadrature's expectation value with $\phi$. The operator $\hat{X}$ is optimal and achieves the maximal sensitivity~(\ref{eq:maxbound}) if its coefficients are chosen as $\mathbf{c}=\mathbf{c}_{\rm{opt}}$, with
\begin{align}\label{eq:copt}
	\mathbf{c}_{\rm{opt}}=\gamma \boldsymbol{\Gamma}^{-1}\mathbf{D},
\end{align}
where $\gamma\in\mathbb{R}$ is a normalization constant.
%\begin{widetext}
	The elements of $\boldsymbol{\Gamma}$ and $\mathbf{D}$ for the ideal state of Eq.(1) of the main paper can be straightforwardly derived using canonical commutation relations. 
	We introduce the quadrature observables $\hat{q}_i(\varphi)=\frac{1}{2}(\hat{a}_i e^{-i\varphi}+\hat{a}_i^{\dagger} e^{i\varphi})$, such that $\hat{q}_i(0)=\hat{x}_i$ and $\hat{q}_i(\pi/2)=\hat{p}_i$. We obtain for the first moments
	\begin{align}
		2\mathcal{N}\langle\hat{q}_1(\varphi)\rangle&=\alpha_{\varphi}  e^{-i \phi}+\alpha_{\varphi} ^* e^{i \phi }+\left(\alpha_{\varphi}+ \alpha_{\varphi}^*\right) \left(2 \left| \alpha \right| ^2 (\cos (\phi )+1)+3\right),
	\end{align}
	where we introduced $\alpha_{\varphi}=\alpha e^{-i\varphi}$, and for the second moments
	\begin{align}
		2\mathcal{N}\langle\hat{q}_1(\varphi_1)\hat{q}_1(\varphi_2)\rangle= & \alpha_{\varphi_1}\alpha_{\varphi_2} \left(\left| \alpha \right| ^2 (\cos (\phi )+1)+e^{-i \phi }+2\right)+  \alpha_{\varphi_1}^*\alpha_{\varphi_2} ^*\left(\left| \alpha \right| ^2 (\cos (\phi )+1)+e^{i \phi }+2\right) \notag \\
		& +\cos \left(\varphi _1-\varphi _2\right) \left(2 \left| \alpha \right| ^4 (\cos (\phi )+1)+2 |\alpha|^2 (\cos (\phi)+2)+1\right) \\
		& +e^{-i \left(\varphi _1-\varphi _2\right)} \left(|\alpha|^2 (\cos (\phi )+1)+1\right),\notag
	\end{align}
	where $\mathcal{N}=2 (1+\left| \alpha \right|^2  (\cos(\phi )+1))$ is the state's normalization factor. The same expressions with $\phi$ replaced by $-\phi$ describe the expectation values for the corresponding observables in the second mode. With these expressions we are able to construct the vector $\mathbf{D}$ and the diagonal $2\times 2$ blocks of the $4\times 4$ covariance matrix $\boldsymbol{\Gamma}$. The off-diagonal blocks are obtained with
	\begin{align}
		2\mathcal{N}\langle\hat{q}_1(\varphi_1)\hat{q}_2(\varphi_2)\rangle=&4 \left| \alpha \right| ^2 \cos \left(\varphi _1-\varphi _2\right) \notag \\
		& +(2 \left| \alpha \right| ^2+1) \cos \left(\varphi _1-\varphi_2+\phi \right)+ \left| \alpha \right| ^2 (\cos (\phi )+1) \left(\alpha_{\varphi_1}+\alpha_{\varphi_1}^*\right)\left(\alpha_{\varphi _2}+\alpha_{\varphi_2}^*\right)\notag\\
		& +(\cos (\phi )+2) \left(\alpha_{\varphi _1}\alpha_{\varphi _2}+\alpha^*_{\varphi _1}\alpha^*_{\varphi _2}\right).
	\end{align}
	The sensitivity~(\ref{eq:maxbound}) is maximized at $\phi=\pi$, where we find
	\begin{align}\label{eq:idealSupp}
		\max_{\mathbf{c}}\frac{\left|\partial_{\phi}\langle \hat{X}\rangle\right|^2}{(\Delta\hat{X})^2}=\frac{2}{3}|\alpha|^2,
	\end{align}
	with the optimal observable given according to~(\ref{eq:copt}) as $\mathbf{c}_{\mathrm{opt}}=(-\mathrm{Im}(\alpha),\mathrm{Re}(\alpha),\mathrm{Im}(\alpha),-\mathrm{Re}(\alpha))^T/|\alpha|$, up to an arbitrary normalization factor.
	
	An unconstrained maximization over all possible observables (not limiting to homodyne measurements) yields the ultimate quantum sensitivity limit~\cite{GessnerPRL2019}
	\begin{align}
		\max_{\hat{X}}\frac{\left|\partial_{\phi}\langle \hat{X}\rangle\right|^2}{(\Delta\hat{X})^2}=F_Q[\Psi(\phi)],
	\end{align}
	where $F_Q[\Psi(\phi)]$ is the quantum Fisher information~\cite{BraunsteinCavesPRL1994}. It can be obtained as~\cite{PezzeSmerzi2014}
	\begin{align}\label{eq:qfipure}
		F_Q[\Psi(\phi)]=4\left(\langle \partial_{\phi}\Psi(\phi)|\partial_{\phi}\Psi(\phi)\rangle-\left|\langle \partial_{\phi}\Psi(\phi)|\Psi(\phi)\rangle\right|^2\right),
	\end{align}
	leading to
	\begin{align}
		F_Q[\Psi(\phi)]=\frac{2|\alpha|^2+1}{(1+|\alpha|^2(1+\cos(\phi)))^2}.
	\end{align}
	The maximum sensitivity $F_Q[\Psi(\pi)]=2|\alpha|^2+1$ is attained at $\phi=\pi$.
%\end{widetext}

\subsection{Experimental imperfections}
All the imperfections of the experimental setup can be taken into account by employing two parameters: the preparation efficiency $\eta_p$ and the detection efficiency $\eta_d$.

The first parameter, $\eta_p$, considers the purity of the added single photon and the unwanted counts in the single-photon detector. The heralded state is thus a mixture of the ideal state, with weight $\eta_p$, with the input uncorrelated two-mode coherent state of amplitude $\alpha$.
This parameter mainly depends on the spatio-temporal width of the filters and on the dark counts of the single-photon detector. In the case of $\alpha=0$, the preparation efficiency for single-photon states is $\eta_{p\text{SP}}=0.92$.
However, since the Mach-Zehnder interferometer has a nonperfect visibility ($V=99.6\%$), it introduces unwanted counts whose rate depends on the seed amplitude. As a consequence, the preparation efficiency decreases with $\alpha$.
By performing independent measurements of the spurious trigger events as a function of the seed amplitude, with the interferometer phase $\phi$ set to $\pi$, we find an empirical relation connecting $\eta_p$ to $\alpha$: $\eta_p(\alpha)=\eta_{p\text{SP}} \: / (0.0052\: \alpha^2 + 1)$ for the relevant case of real $\alpha$.

The second parameter, $\eta_d$, considers optical losses, quantum efficiency of photodiodes, electric noise in the homodyne detector and the mode-matching between signal and LO. According to the standard quantum model of losses \cite{Leonhardt97}, they can be modeled by attenuating each mode of the generated state by a beam-splitter with transmission $\eta_d$.
The detection efficiency is evaluated before the measurements, by fitting the quadrature distribution of single photon states and considering a preparation efficiency $\eta_{p\text{SP}}=0.92$. We obtain a mean value $\bar{\eta}_d=59\%$.\\
%\begin{widetext}
	All these effects considered, we derived the analytical expressions for the mean value and the variance of the operator $\hat{X}=\hat{p}_1-\hat{p}_2$:
	\begin{equation}
		\expval*{\hat{X}}=-\frac{\alpha \eta_p \sqrt{\eta_d}  \sin (\phi )}{1+ (1+\cos (\phi ))\alpha ^2},
		\label{eq:meanXeff}
	\end{equation}
	\begin{equation}
		(\Delta \hat{X})^2=\frac{\cos (\phi ) \left(\alpha ^2-\eta_d \eta_p\right)+\alpha ^2+\eta_d \eta_p+1}{2 \left(1+ (1+\cos (\phi ))\alpha ^2\right)}-\frac{\alpha ^2 \eta_d \eta_p^2 \sin ^2(\phi )}{\left(1+ (1+\cos (\phi ))\alpha ^2\right)^2}.
		\label{eq:varXeff}
	\end{equation}
	At $\phi=\pi$ we obtain the sensitivity
	\begin{equation}
		S=\frac{\left|\partial_{\phi}\langle \hat{X}\rangle\right|^2}{(\Delta\hat{X})^2}=\frac{2\alpha^2\eta_d\eta_p^2}{1+2\eta_d\eta_p},
	\end{equation}
	which converges to the ideal result~(\ref{eq:idealSupp}) in the limit $\eta_d,\eta_p\to 1$. Starting from these equations we derived the theoretical predictions presented as dashed red lines in Figures (3-5) of the main paper. It is also worth noting that, taking into account the experimental imperfections, the choice of the coefficients $\mathbf{c}=(c_{x1},c_{p1},c_{x2},c_{p2})^T $ to determine the optimal observable $\hat{X}$ does not change.
%\end{widetext}
\end{widetext}

\end{document}